\documentclass[aps,twocolumn,showpacs,superscriptaddress,prb,10pt,nofootinbib]{revtex4-2}
\usepackage{amsmath}  % needed for \tfrac, \bmatrix, etc.
\usepackage{amsfonts} % needed for bold Greek, Fraktur, and blackboard bold
\usepackage{graphicx} % needed for figures

\usepackage{xcolor} 

\usepackage{tikz}
\usepackage{pgfplots}
\usepackage{placeins}

\usepackage[normalem]{ulem}

\begin{document}

\title{Tuning Interatomic Forces with Magnetic Fields: Feshbach Resonances in Lithium-6}

\author{Ettore Vitali}
\affiliation{Department of Physics, California State University, Fresno, Fresno, California 93740, USA}

\author{Gino Edward Gamboni}
\affiliation{Department of Physics, California State University, Fresno, Fresno, California 93740, USA}
\affiliation{Department of Physics, University of California, Merced, California 95343, USA}
\date{\today}

\begin{abstract}
Feshbach resonances, first studied in the context of nuclear reactions, have since become a cornerstone of modern atomic physics. They offer a remarkable degree of control over interatomic (and even intermolecular) interactions by tuning external magnetic fields. This tunability arises from the interplay between quantum scattering and the internal structure of atoms or nuclei. In this work, we explore the essential physics of Feshbach resonances using only basic quantum mechanics, aiming to make this powerful concept accessible to educators and students alike.
\end{abstract}
% AJP requires an abstract for all regular article submissions.
% Abstracts are optional for submissions to the "Notes and Discussions" section.

\maketitle % title page is now complete

\section{Introduction} % Section titles are automatically converted to all-caps.
% Section numbering is automatic.

Cooling a gas of atoms to just a few billionths of a degree above absolute zero brings the system into a regime where quantum effects dominate its behavior. In this ultracold domain, {\it{Feshbach resonances}} \cite{FESHBACH1958357,RevModPhys.82.1225} play a central role: by adjusting an external magnetic field, one can tune the interatomic forces from strongly attractive to strongly repulsive—or even make them vanish altogether. Ultracold gases thus provide an exceptionally controllable platform for exploring the principles of quantum many-body physics \cite{RevModPhys.80.885}.

Although experiments involving Feshbach resonances represent the frontier of modern atomic physics \cite{RevModPhys.82.1225}, the essential ideas behind them can be understood using only basic quantum mechanics \footnote{Readers interested in a simpler one-dimensional presentation that still captures some qualitative features of the Feshbach resonance may consult Ref.~\cite{taron2013}.}. In this paper, we explore a central and intuitive question: {\it{what is the mechanism by which an external magnetic field controls the interaction between two $^6$Li atoms?}}
This question can be addressed using tools accessible to advanced undergraduate students and provides a concrete example of how fundamental quantum concepts manifest in real physical systems. We focus on $^6$Li atoms because they are widely used in ultracold-atom experiments involving fermionic quantum gases, together with $^{40}$K \cite{doi:10.1126/science.1079107,PhysRevLett.91.080406}.

To guide the reader through the key ideas, we begin in Sec.~\ref{Sec:BasicScatt} with a familiar Hamiltonian and review central results from scattering theory, emphasizing phase shifts and the scattering length. In Sec.~\ref{Sec:squarewell}, we use the example of a spherical square-well potential to illustrate the emergence of resonances. We then discuss the physics of Feshbach resonances and their realization in $^6$Li in Secs.~\ref{Sec:feshbach} and \ref{Sec:twochannel}. Finally, Sec.~\ref{Sec:conclusion} summarizes the main insights and their pedagogical implications.

\section{A few important notions from basic scattering theory}
\label{Sec:BasicScatt}
Before we dive into the subtleties of $^6$Li collisions, we find it useful to review a few important points from scattering theory. 
%This \add{allows the teacher a smooth} transition from the study of a single-particle\add{'s Hamiltonian} in a central potential (e.g., the hydrogen atom) into the more advanced nuances of scattering phenomena. % involving indistinguishable particles with internal states. 
We start from a Hamiltonian of the form
\begin{equation}
    \hat{H} = - \frac{\hbar^2}{2m} \nabla^2 + U(r),
    \label{BasicScatt:ham}
\end{equation}
where $\hbar$ is the reduced Planck constant and $\nabla^2$ is the standard Laplacian operator. We interpret \eqref{BasicScatt:ham} as describing the relative motion of two particles with a reduced mass $m$ interacting through the interatomic potential $U(r)$, which only depends on the separation $r$ between the particles.
%The hamiltonian \eqref{BasicScatt:ham} may describe either the motion of one particle with mass $m$ in a central external field or the relative motion of two particles interacting through the interatomic potential $U(r)$ which depends only on the distance $r$ between the particles. In the latter case, which will be our main concern, $m$ is the reduced mass of the two particles. 
In our discussion, we shall consider potentials that fall to zero outside some range $b$, i.e. $U(r)\to 0$ if $r\gg b$.
%incidentally, we comment that, for cold atomic systems, $b$ is typically much shorter than the average inter-particle distance.

We focus on solutions of the time-independent Schr\"odinger equation with positive energies, usually referred to as {\it{scattering states}}. Since $E>0$, we can always introduce a parameter $p$ such that $E \equiv p^2/(2m)$. At this stage, $p$ is simply a convenient label for the states. As we will see below, in the asymptotic region $r\gg b$, where the interaction potential becomes negligible, $p$ acquires the physical meaning of the relative momentum between the two particles.
%We focus on solutions of the time-independent Schr\"odinger equation with positive energies $E \buildrel{\text{def}}\over{=}  \frac{p^2}{2m}$. Such solutions are usually referred to as {\it{scattering states}}. %\add{We shall make the substitution $E \buildrel{\text{def}}\over{=} \frac{p^2}{2m}$ as the equations used later on in this article are more suitably expressed in terms of momentum $p$ instead of the energy $E$.} 
Leveraging the conservation of angular momentum in the same way as we do when we study the hydrogen atom, it can be shown \cite{griffiths} that, for a given value of $E$ (or $p$), every scattering state $\phi_{p}(\vec{r})$ can be written as a linear combination of basis functions. That is, we have
\begin{equation}
    \phi_{p}(\vec{r}) = \sum_{l=0}^{+\infty} \sum_{m=-l}^{l} c_{p,l,m} \, \frac{y_{p,l}(r)}{r} \, Y_{l,m}(\theta,\varphi),
    \label{BasicScatt:lincomb}
\end{equation}
where $(r,\theta,\varphi)$ are the spherical coordinates of the vector $\vec{r}$, $c_{p,l,m} \in \mathbb{C}$ are coefficients and $Y_{l,m}(\theta,\varphi)$ are the spherical harmonics \cite{griffiths} which are eigenvectors of $\hat{L}^2$ and $\hat{L}_z$, $\vec{\hat{L}}$ being the angular momentum. The function $y_{p,l}(r)$ is found by solving 
 the radial equation \cite{griffiths}
\begin{equation}
 - \frac{\hbar^2}{2m} \frac{d^2 y_{pl}(r)}{dr^2} + \left(U(r)  + \frac{\hbar^2 l (l+1)}{2mr^2}   \right) \, y_{pl}(r) = \frac{p^2}{2m} \, y_{pl}(r) 
 \label{BasicScatt:rel}
\end{equation}
with the condition $y_{pl}(0)=0$. 

We observe that, while the behavior of $y_{pl}(r)$ for small $r$, say $r \lesssim b$,
can be complicated and strongly depends on the details of the function $U(r)$, in the asymptotic region, i.e., when $r\to +\infty$, both the interatomic potential and the centrifugal term $\frac{\hbar^2 l (l+1)}{2mr^2}$ become negligible, leaving us with what appears mathematically as a one-dimensional Schrödinger equation for a free particle.
This implies that the solution in the asymptotic region can be written as
\begin{equation}
y_{p,l}(r) \buildrel{r \to +\infty}\over{\to} \sqrt{\frac{2}{\pi \hbar}} \sin\left(\frac{pr}{\hbar} - \frac{\pi}{2}l + \delta_l(p) \right) ,
\label{BasicScatt:phaseshiftdef}
\end{equation}
where $\delta_l(p)$ is the {\it{phase shift}} and encapsulates the effects of $U(r)$. 
We also note that $\delta_l(p)=0$ if the particles do not interact.

\begin{figure}[h!]
    \centering
    \includegraphics[width=0.7\linewidth]{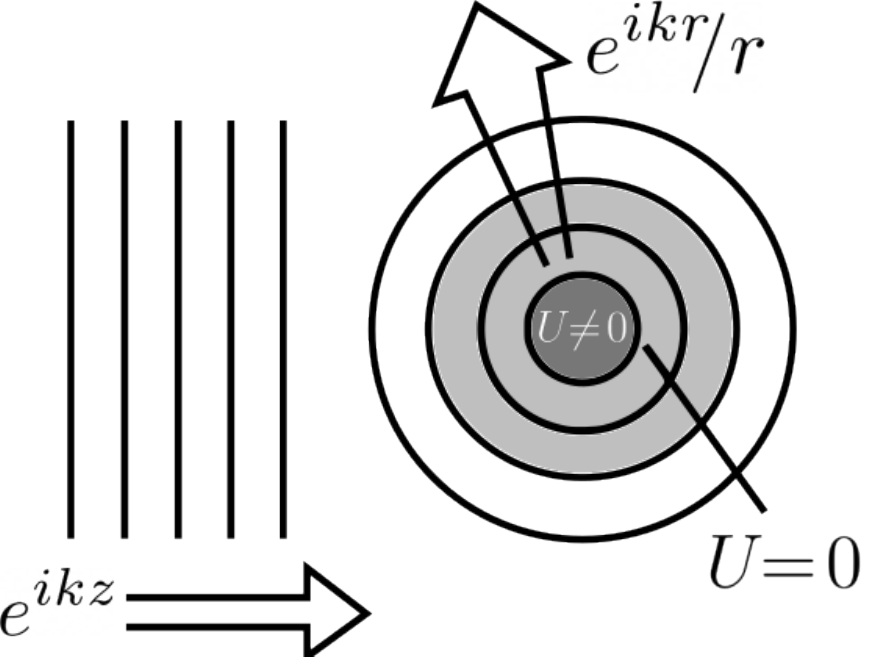}
    \caption{Pictorial representation of Eq.~\eqref{BasicScatt:uplusfar}: a plane wave ($e^{i k z}$) propagates toward the scattering center along the $z$ axis, and a spherical wave ($e^{i k r}/r$) emerges from it.}
    \label{fig: r to z wave}
\end{figure}

%while it is very convenient from a mathematical point of view to leverage spherical symmetry to be able to simplify the Schr\"odinger equation through the separation of variables, which leads to the states $u_{p,l,m}(\vec{r})$, 
To connect the expansion~\eqref{BasicScatt:lincomb} with the physical picture of scattering shown in Fig.~\ref{fig: r to z wave}, namely an incoming plane wave interacting with the potential and producing an outgoing spherical wave, we construct a new basis in the energy eigenspace parametrized by a vector $\vec{p}$, with magnitude $|\vec{p}| = p$.
We denote the polar angles defining the direction of $\vec{p}$ as $(\alpha,\beta)$ and choose the coefficients in \eqref{BasicScatt:lincomb} as
\begin{equation}
c_{p,l,m} \buildrel{\text{def}}\over{=} e^{i \delta_l(p)} \frac{i^l}{p} Y^{\star}_{lm}(\alpha,\beta).
\end{equation}
This choice implies the following asymptotic behavior of the new basis functions $u_{\vec{p}}(\vec{r})$ \cite{joachain1975quantum}
\begin{equation}
u_{\vec{p}}(\vec{r}) \buildrel{r \to +\infty}\over{\to} \frac{1}{(2\pi \hbar)^{3/2}} \left( e^{\frac{i}{\hbar} \vec{p} \cdot \vec{r}} + \frac{e^{i \frac{p r}{\hbar}}}{r} f(\theta,\varphi; \vec{p}) \right),
\label{BasicScatt:uplusfar}
\end{equation}
which follows from combining \eqref{BasicScatt:phaseshiftdef} with the well-known expansion of a plane wave as a linear combination of spherical waves \cite{joachain1975quantum,landau_lifshitz_qm}. Here, the {\it scattering amplitude} is given by \cite{joachain1975quantum}
\begin{equation}
f(\theta,\varphi; \vec{p}) = \frac{\hbar}{p} \sum_{l=0}^{+\infty}
(2l + 1)\, e^{i \delta_l(p)} \sin \delta_l(p)\,
P_l(\cos \theta_{\hat{p r}}),
\label{BasicScatt:fscatamp}
\end{equation}
where $\theta_{\hat{p r}}$ is the angle between $\vec{r}$ and $\vec{p}$, and $P_l$ is the Legendre polynomial of order $l$.
Equation \eqref{BasicScatt:uplusfar} describes a superposition of a plane wave—propagating in the direction $\vec{p}/p$ with wavelength $2\pi \hbar/p$—and an outgoing spherical wave (since $p>0$) of the same wavelength. Figure~\ref{fig: r to z wave} provides a pictorial representation of these waves.

Equation \eqref{BasicScatt:uplusfar} helps us build a physical picture of the collision: at times long before the collision, the two particles approach each other with well-defined momentum $\vec{p}$. After the collision, the system’s state becomes a superposition of a plane wave continuing along the direction $\vec{p}/p$, and a scattered spherical wave emerging from the collision.

While \eqref{BasicScatt:uplusfar} provides the mathematical foundation for this picture, it does not include explicit time dependence. This may prompt readers to question the meaning of "before" and "after" the collision. To rigorously describe the time-dependent scenario where particles travel toward each other and collide producing a scattered wave, we need to construct wave packets as superpositions of the basis functions $u_{\vec{p}}(\vec{r})$
\begin{equation}
\psi(\vec{r}, t) = \int d^3 p \, c(\vec{p}) \, u_{\vec{p}}(\vec{r}) \, e^{-\frac{i}{\hbar} \frac{p^2}{2m} t},
\label{BasicScatt:wavepacket}
\end{equation}
where the coefficient function $c(\vec{p})$ is significant only near a particular momentum vector $\vec{p}_0$. Details of the derivations are provided in the supplementary material; here we summarize the key results relevant to scattering amplitudes and lengths, which are the key ingredients we need to study resonances.

In the asymptotic region, Eqs.~\eqref{BasicScatt:uplusfar} and \eqref{BasicScatt:wavepacket} allow us to write
$\psi(\vec{r}, t) \buildrel{r \to +\infty}\over{\to} \psi_{\text{in}}(\vec{r}, t) + \psi_{\text{scatt}}(\vec{r}, t)$.
Here, the {\it incoming wave} $\psi_{\text{in}}(\vec{r}, t)$ represents the relative motion of the two particles along the classical trajectory $\vec{r} = \frac{\vec{p}_0}{m} t$, where $\vec r$ is the relative coordinate and $m$ the reduced mass, with the wave function being non-negligible only in a region consistent with quantum uncertainties.

The {\it scattering wave function}
\begin{equation}
\psi_{\text{scatt}}(\vec{r}, t) = \frac{1}{(2\pi \hbar)^{3/2}} \int d^3 p \, c(\vec{p}) \, \frac{e^{i \frac{p}{\hbar} r - \frac{i}{\hbar} \frac{p^2}{2m} t}}{r} f(\theta, \varphi; \vec{p}),
\label{BasicScatt:psiscat}
\end{equation}
encodes the effects of the collision. Importantly, $\psi_{\text{scatt}}(\vec{r}, t)$ vanishes as $t \to -\infty$, confirming our intuitive picture: {\it before} the collision, formally at $t \to -\infty$, the particles do not interact and move toward each other with relative momentum $\vec{p}_0$.
In addition, \eqref{BasicScatt:psiscat} represents a spherical wave packet moving radially outward with velocity $p_0/m$.
{\it After} the collision ($t \to +\infty$), in any direction $(\theta, \varphi)$ different from that of $\vec{p}_0 / p_0$, only the scattered spherical wave remains, with an amplitude controlled by the factor $f(\theta, \varphi; \vec{p})$.

%It appears natural then to expect that the scattering amplitude $|f(\theta, \varphi; \vec{p})|^2$ gives the probability of the incoming particle scattering into the solid angle around $(\theta, \varphi)$.
Incidentally, we mention that it can be proved, by considering a monochromatic particle beam impinging on a target, that $|f(\theta, \varphi; \vec{p})|^2$ coincides with the differential cross-section, which quantifies the fraction of particles scattered into a given solid angle per unit solid angle. For further details, see \cite{joachain1975quantum}.

The above discussion hopefully sheds light into the crucial point that, so long as we observe the particles when they are ``far enough'' from each other, all the information about the interaction is encapsulated in the scattering amplitude $f(\theta,\varphi; \vec{p})$. In general, we need to sum over all possible values of $l$ and $m$ 
but in order to emphasize physical interpretation in a simple setting, in this work we focus on the $l=0$ term, the so-called {\it{$s$-wave}}, which is also the most relevant for low energy collisions \cite{joachain1975quantum}; simple manipulations give us
\begin{equation}
  f_{l\,=\,0}(\theta,\varphi; \vec{p}) =  - \frac{i \hbar}{2 p } \,
\left( e^{2i \delta_0(p)} - 1 \right) = \frac{\hbar}{p} \frac{ 1 }{  \cot \delta_0(p)  - i  }
% , \quad |f_{l=0}(\theta,\varphi; \vec{p})|^2 = \frac{\hbar^2}{p^2} \frac{ 1 }{  \cot^2 \delta_0(p)  + 1  }
    \label{BasicScatt:fscatamps}
\end{equation}
%where the last step is allowed only if $\delta_0(p) \neq n \pi$, $n \in \mathbb{Z}$. 
Notice that the $s$-wave scattering amplitude does not depend on the direction of $\vec{p}$ or on the angles $(\theta,\varphi)$. This implies that $s$-wave scattering generates an isotropic outgoing spherical wave. A moment of reflection will suggest that large values of $|f_{l\,=\,0}(\theta,\varphi; \vec{p})|^2$ correspond to ``strong'' interactions (higher chance of observing a scattered wave)
%, while small values correspond to ``weak'' interactions, 
and the key contributing factor is the dependence of the phase shift $\delta_0$ on $p$ (or $E$). 
When we study low-energy collisions, it is very useful to define a characteristic length scale, called the $s$-wave {\it{scattering length}}, which encapsulates the low-energy effects of the interaction potential
\begin{equation}
    a_0 \buildrel{\text{def}}\over{=} - \hbar \lim_{p \to 0} \frac{\tan\delta_0(p)}{p} \, .
    \label{BasicScatt:scatlength}
\end{equation}
The quantity $a_0$ has dimensions of length and determines the low-energy behavior of the scattered wave. As we will illustrate below using the square-well potential, $a_0$ provides a characteristic length scale associated with low-energy scattering and offers a convenient way to characterize the interaction strength.
In practice, $a_0$ is the only quantity we need when we wish to study a low-energy $s$-wave collision. Moreover, different potentials $U(r)$ corresponding to the same value of $a_0$ will have the same effects. This is a tremendous advantage as it allows us to forget about the details of $U(r)$, which can be very complicated, and just focus on $a_0$.

In summary, the details of the interaction potential $U(r)$ enter low-energy scattering only through the scattering length $a_0$. For the square-well potential discussed below, $a_0$ depends on the dimensionless parameter $x=b\hbar^{-1}\sqrt{2m|U_0|}$, which combines the range $b$ and strength $U_0$ of the interaction.

%{\color{green} The definition \eqref{BasicScatt:scatlength} is not intuitive in relation to the actual physical meaning of $a_0$. We will discuss below the example square-well potential, which will help us illuminate the physical meaning and the importance of the scattering length. More details are given in the supplementary material. }
%Simple manipulations also give us
%\begin{equation}
%   $ |f_{l=0}(\theta,\varphi; \vec{p})|^2 = \frac{\hbar^2}{p^2} \frac{ 1 }{  \cot^2 \left( -\frac{p}{\hbar} b \right)  + 1  } \buildrel{pb/\hbar \ll 1}\over{\approx} b^2. $
%\end{equation}

\section{The finite spherical square-well potential and zero-energy resonances}
\label{Sec:squarewell}
\begin{figure}[t]  % 't' means top of the column; you can also use [h] or [b]
    \centering
    \includegraphics[width=\columnwidth]{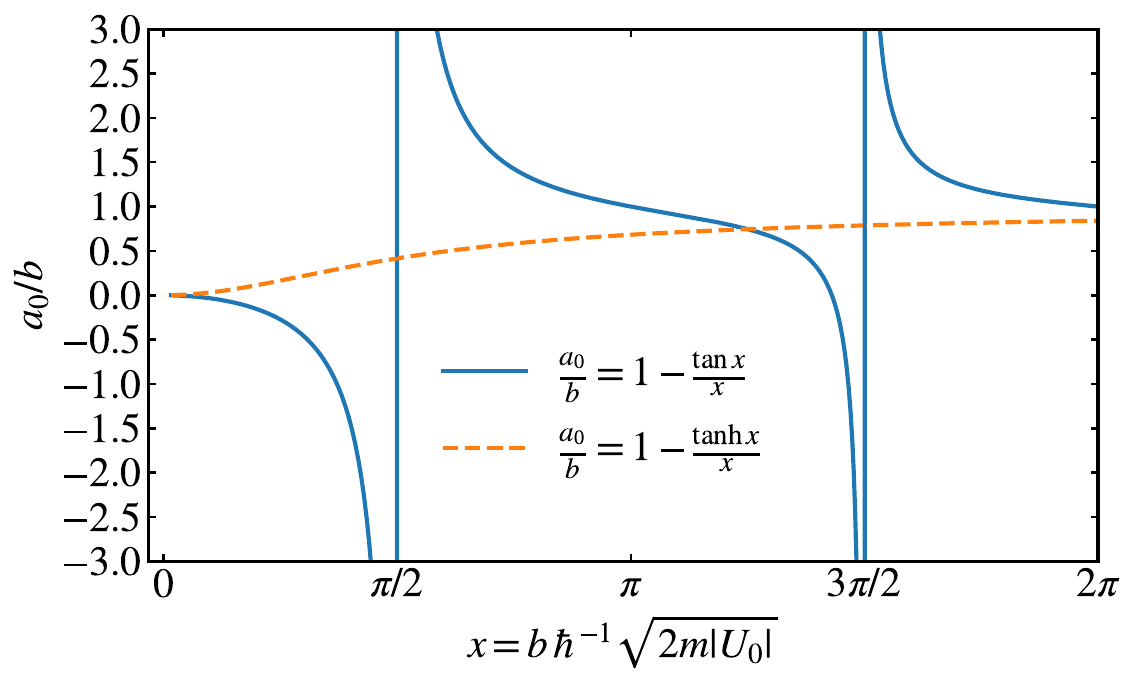}
    \caption{(Color online) Graph of the square-well scattering length relation~\eqref{BasicScatt:scatlengthsqwell}, $a_0/b=1-(\tan{x})/x$ (solid line) for the attractive case and $a_0/b=1-(\tanh{x})/x$ (dashed line) for the repulsive case, where $x=b\hbar^{-1}\sqrt{2m|U_0|}$. The vertical lines indicate the values of $x$ for which the scattering length diverges in the attractive case.}
    \label{fig: a_0(x)/b plot}
\end{figure}

Let us consider a square-well attractive potential $U(r)=-U_0$ if $r < b$ and $U(r) = 0$ if $r \geq b$, with $U_0 > 0$. The analysis of the $s$-wave radial equation is done in the supplementary material and leads to the following result
\begin{equation}
    a_0 = b \left(1 - \frac{1}{\frac{\sqrt{2 m U_0}}{\hbar} \, b} \tan\left(\sqrt{2 m U_0} \frac{b}{\hbar}  \right)  \right) \, ,
    \label{BasicScatt:scatlengthsqwell}
\end{equation}
which explicitly shows that $a_0$ carries information about both the range and the strength of the interaction.
We incidentally observe that this expression can also be used in the repulsive case $U_0 < 0$ using the identity $\tan\left(i\sqrt{2 m |U_0|} \frac{b}{\hbar}  \right) = i \tanh\left( \sqrt{2 m |U_0|} \frac{b}{\hbar}  \right)$. 
In Fig.~\ref{fig: a_0(x)/b plot} we show $a_0/b$ as a function of
the dimensionless parameter $x \buildrel{\text{def}}\over{=} b\hbar^{-1}\sqrt{2m|U_0|}$ for both the attractive and repulsive cases.

In the repulsive case the scattering length is always positive and it interpolates smoothly between non-interacting behavior $a_0\approx 0$ if $x \ll 1$ and hard-sphere behavior $a_0 \approx b$ if $x \gg 1$ (see the supplementary material for details). 
The attractive case is more nuanced due to the highly nontrivial behavior of the right-hand side of \eqref{BasicScatt:scatlengthsqwell}. For small values of $x$ the scattering length is negative, suggesting that the sign of $a_0$ contains information about whether the interaction is repulsive or attractive. When $x = \frac{\pi}{2}(2n+1)$, $n=0,1,2,\dots$, the scattering length diverges. Such divergences have a profound impact on the low-energy scattering amplitude discussed in Eq.~\eqref{BasicScatt:fscatamps}. Whenever $a_0$ is finite, Eq.~\eqref{BasicScatt:scatlength} guarantees that the scattering amplitude remains finite as $p\to0$. However, when $a_0$ diverges, the low-energy scattering becomes resonantly enhanced, corresponding to $\delta_0(p)\to\pi/2$ in the limit $p\to0$.

The divergence of the scattering length is directly connected to the appearance of a zero-energy bound state in the square well, i.e., a solution of the Schr\"odinger equation with $E=0$.
When $l=0$, the solution of the radial equation (i.e., Eq.~\eqref{BasicScatt:rel} with $E$ instead of $p^2/2m$) 
must have the form
\begin{equation}
    y_E(r) = \begin{cases}
        A \sin\left( \hbar^{-1} \sqrt{ 2m (E + U_0) } \, r \right), \quad r < b \\
        C e^{- \hbar^{-1} \sqrt{ 2m |E| } \, r  }, \quad r > b
    \end{cases}
    \, ,
\end{equation}
where $A$ and $C$ are coefficients. Both $y_E(r)$ and $\frac{dy_E(r)}{dr}$ must be continuous at $r=b$, which leads to the following non-linear equation for the energy eigenvalues
\begin{equation}
\begin{split}
    &       \sqrt{ (E + U_0) } \cot\left( \hbar^{-1} \sqrt{ 2m (E + U_0) } \, b \right) = - \sqrt{  |E| } \, .
\end{split}
\end{equation}
We notice in particular that this equation has the solution $E=0$ if and only if $x=b \hbar^{-1}\sqrt{2m U_0} = \frac{\pi}{2}(n+1)$
which is exactly the condition in which the scattering length diverges.

At the threshold, the state emerges at zero energy and, as \(x\) increases further beyond this point—that is, as the potential well becomes deeper or wider—its energy becomes negative.

The appearance of bound states and the resulting divergence of the scattering length \(a_0\) have a deep impact on how particles scatter. The crucial point is how the phase shift \(\delta_0(p)\) changes with \(p\), and how this variation affects the formation of a scattered wave packet by integrating over $\vec{p}$ in Eq.~\eqref{BasicScatt:psiscat}.
Imagine a wave packet centered around some \(p_0\). The scattered wave depends on how the phase shifts vary close to that momentum (see Eq.~\eqref{BasicScatt:fscatamps}). 
When the magnitude of $a_0$ is large, even small differences in momentum cause big differences in phase shifts.
We have seen that the scattered wave packet propagates outward with the same group velocity 
$v = p_0/m$ as a free particle. However, the interaction introduces an important twist: 
the wave packet experiences a \emph{time delay}~\cite{fetic2024wigner}. 
In simple terms, when $a_0 < 0$, the outgoing packet emerges 
\emph{later} than it would in the absence of any potential. This delay, which reflects 
the extra time the particles spend in the interaction region, can be quantified 
(see Supplementary Material) as 
\begin{equation}
t_{\mathrm{delay}}
=
2\hbar\,\frac{m}{p_0}\,\frac{d\delta_0(p_0)}{dp}
\approx
-2\,\frac{m}{p_0}\,a_0 .
\end{equation}
Near a zero-energy resonance, where $a_0$ becomes large and negative, the delay grows 
substantially—the particles remain close together for a long time before separating. 
In addition, the scattered wave packet acquires a much larger amplitude, reflecting the strong enhancement of low-energy scattering near resonance.

When $a_0$ changes sign and becomes positive, $t_{\mathrm{delay}}$ also changes sign, 
and the dynamics appear as if the interaction were effectively \emph{repulsive}. 

\begin{figure}[t]  % 't' means top of the column; you can also use [h] or [b]
    \centering
    \includegraphics[width=\columnwidth]{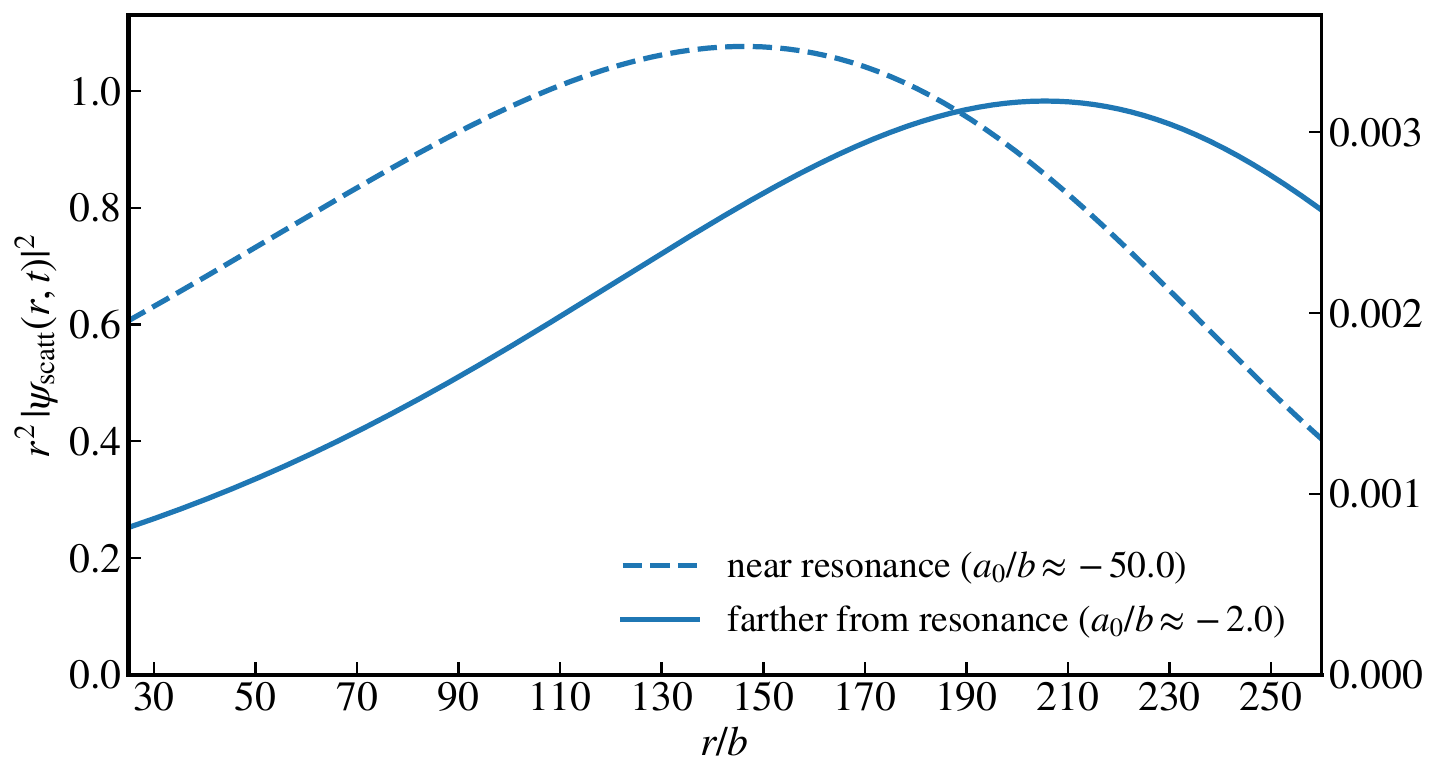}
    \caption{Snapshots of two outgoing $s$-wave scattering wave packets 
$\psi_{\mathrm{scatt}}(r,t)$ at the same time $t\,\frac{\hbar}{2 m b^{2}} = 10^{4} > 0$, shown as functions 
of separation $r$ in the asymptotic region $r \gg b$. 
Both packets are constructed using the same mean momentum $p_{0} b/\hbar = 0.015$ (low energy) and 
momentum width $\Delta p b / \hbar \simeq 0.03$, but different depths of the attractive 
square well. 
For the dashed curve, the potential is tuned close to resonance, 
giving a large negative scattering length $a_0/b \simeq -50$, 
while the solid curve corresponds to a case farther from resonance, with 
$a_0/b \simeq -2$. 
The two vertical axes reflect the respective physical amplitudes of the 
radial probability density for both wave packets; the right (unlabeled) axis applies to the solid curve 
and is scaled so that both traces can be clearly compared in a single plot.}
    \label{fig:delay-time-plot}
\end{figure}

Figure~\ref{fig:delay-time-plot} illustrates this behavior by comparing two outgoing 
$s$-wave packets at the same time $t>0$ and mean momentum $p_0$, but with different 
(negative) scattering lengths. The dashed curve, corresponding to a potential tuned 
near resonance ($a_0/b \simeq -50$), is clearly shifted and delayed relative to the 
solid curve ($a_0/b \simeq -2$), which represents a case farther from resonance. 
For positive $a_0$, the shift would reverse direction, signaling the transition to 
effective repulsion.

\section{Collisions between $^6$Li atoms and Feshbach resonances}
\label{Sec:feshbach}

We are now ready to discuss the more subtle case of collisions between two atoms that have internal (hyperfine) degrees of freedom. 
Here we focus on two $^6$Li atoms, although the formalism applies to any system of identical fermions. 

A fully microscopic treatment would require solving the complete electronic Hamiltonian for the six electrons and two nuclei of the atomic pair, including relativistic and spin--orbit terms. 
To keep the analysis tractable while retaining the essential physics, we consider an effective Hamiltonian for the two-atom system of the form
\begin{equation}
    \hat{H} = - \frac{\hbar^2}{2m} \nabla^2 
    + \hat{H}^{\text{atomic}}_1 
    + \hat{H}^{\text{atomic}}_2 
    + \hat{V},
    \label{lithium:hamiltonian}
\end{equation}
where the first term represents the kinetic energy associated with the relative motion of the two atomic centers of mass, 
$\hat{H}^{\text{atomic}}_i$ describes the internal structure of atom $i$, 
and $\hat{V}$ is the interatomic interaction potential. 
We assume that $\hat{V}$ is spherically symmetric and depends on the internal states of the atoms; 
in general, the interaction can depend on the spin configuration and couple different internal states.

We use Dirac notation to denote the internal eigenstates of a single atom,
\begin{equation}
\hat{H}^{\text{atomic}} \, |\alpha\rangle = \varepsilon_{\alpha} \, |\alpha\rangle ,
\end{equation}
where we omit the atomic label $i$, since both atoms have the same set of internal states $\{|\alpha\rangle\}$, each defined with respect to its own nucleus.  
For $^6$Li$\, (1s^2 2s)$, the relevant low-energy internal structure arises from the hyperfine coupling between the electronic spin $S=1/2$ and the nuclear spin $I=1$, leading to six hyperfine states~\cite{griffiths1982hyperfine,schlecht1966li_hyperfine}.

The internal energies $\varepsilon_{\alpha}$ and eigenstates $|\alpha\rangle$ depend on the atomic hyperfine structure and can be modified by an external magnetic field through the Zeeman interaction~\cite{eisberg_resnick_1985,griffiths_qm_2018_zeeman}.  
For a given atom in an external magnetic field, there is an additional contribution to $\hat{H}^{\text{atomic}}$ of the form $-\hat{\boldsymbol{\mu}}\!\cdot\!\boldsymbol{B}$, where $\hat{\boldsymbol{\mu}}$ is the total magnetic moment of the atom.  
As a result, both the energies $\varepsilon_{\alpha}(B)$ and the internal spin structure of the states $|\alpha(B)\rangle$ vary with the magnetic field.
In ultracold gases, this tunability of the internal atomic structure is crucial: it allows one to shift the relative energies between different spin channels and thereby control the scattering properties via Feshbach resonances, as will become clear in the two-body analysis below.

The assumption that $\hat{V}$ is spherically symmetric allows us to look for eigenstates of Eq.~\eqref{lithium:hamiltonian} that are also eigenvectors of $\hat{L}^2$ and $\hat{L}_z$ for the relative motion. 
In this section we focus on $s$-wave scattering and restrict our attention to $l=0$.

For a given eigenvalue $E$ of Eq.~\eqref{lithium:hamiltonian}, the expansion~\eqref{BasicScatt:lincomb} can be generalized to include the internal states as
\begin{equation}
    |\Psi_{E} \rangle 
    = \sum_{\alpha>\beta} 
    c_{E,\alpha,\beta}\,|\psi_{E,l=0,\alpha,\beta}\rangle\,|\alpha,\beta\rangle_{A},
\label{lithium:lincomb}
\end{equation}
where we label the single-atom eigenstates $|\alpha\rangle$ ($\alpha=1,\dots,6$) by increasing energy,
$\varepsilon_\alpha > \varepsilon_\beta$ when $\alpha>\beta$, 
and define the antisymmetrized two-atom internal state as 
$|\alpha,\beta\rangle_{A} \equiv \frac{1}{\sqrt{2}}\big(|\alpha\rangle|\beta\rangle - |\beta\rangle|\alpha\rangle\big)$\footnote{The label ``A'' denotes ``antisymmetric.''}.
In addition, the relative-motion wave function has the usual form 
$\langle \vec{r} | \psi_{E,l,\alpha,\beta} \rangle = \frac{y_{E,l,\alpha,\beta}(r)}{r} Y_{l,m}(\theta,\varphi)$\footnote{Notice that we do not write $E = p^2/2m$, as we have not yet restricted $E$ to be positive.} .
The internal states appear in an antisymmetric combination because the atoms are fermions and the spatial part of the wave function is symmetric when $l=0$.

The generalization of the radial equation~\eqref{BasicScatt:rel} for $l=0$ can then be written as~\cite{houbiers1997li_collisions}
\begin{equation}
\begin{split}
& \left( - \frac{\hbar^2}{2 m} \frac{d^2}{dr^2} + \varepsilon_{\alpha} +  \varepsilon_{\beta} - E\right) \, y_{E,\alpha \beta}(r) \\
& = -  \sum_{\alpha^{\prime} \neq \beta^{\prime}} 
V_{\alpha,\beta ; \alpha^{\prime} ,\beta^{\prime}}(r)\, y_{E,\alpha^{\prime} \beta^{\prime}}(r),
\end{split}
\label{lithium:coupled-radial}
\end{equation}
where 
$V_{\alpha,\beta ; \alpha^{\prime} ,\beta^{\prime}}(r) \equiv 
\langle \alpha , \beta |_A \,  \hat{V}(r) \, | \alpha^{\prime} , \beta^{\prime} \rangle_A$.
Each combination $(\alpha,\beta,l)$ defines a \emph{scattering channel}, and Eq.~\eqref{lithium:coupled-radial} constitutes the corresponding set of \emph{coupled-channels} radial equations. 
If $E < \varepsilon_{\alpha} +  \varepsilon_{\beta}$, the channel is \emph{closed}, meaning that the wave function decays exponentially outside the potential range. 
Conversely, if $E > \varepsilon_{\alpha} +  \varepsilon_{\beta}$, the channel is \emph{open}, allowing the two atoms to separate to large distances.

\section{A two-channel square-well model for a Feshbach resonance}
\label{Sec:twochannel}
In the case of $^6$Li atoms at very low collision energies, each atom has six internal hyperfine states associated with the electronic configuration $1s^22s$. 
In principle, this gives rise to multiple coupled channels for $s$-wave scattering, since different spin configurations can interact through the interatomic potential~\cite{houbiers1997li_collisions}. 
However, not all of the $6\times6/2$ antisymmetric combinations contribute to $s$-wave scattering: 
additional selection rules arising from the symmetry of the interatomic (or equivalently, molecular) 
interaction---which depends on the total electronic spin---restrict 
the number of coupled channels to only a few. 

To capture the essential physics without the complexity of a full multi-channel calculation, 
we introduce a simplified two-channel model consisting of one open and one closed channel 
coupled by an off-diagonal interaction. 
In this model, the open channel $|o\rangle$ represents a spin configuration that is energetically accessible at the collision energy, 
while the closed channel $|c\rangle$ has a higher threshold energy. 
For large separations ($r>b$), the wave function in the closed channel decays exponentially, 
whereas within the interaction region ($r<b$) the two channels are coupled, which strongly affects the scattering behavior, as we will now discuss.

The total quantum state of the system can be written as a superposition of components in the two channels,
\begin{equation}
 |\Psi_E \rangle = c_{E,o}\,|\psi_{E,o}\rangle\,|o\rangle 
                 + c_{E,c}\,|\psi_{E,c}\rangle\,|c\rangle,
 \label{lithium:psietwochan}
\end{equation}
where $\langle r|\psi_{E,o}\rangle = y_{E,o}(r)/r$ and 
$\langle r|\psi_{E,c}\rangle = y_{E,c}(r)/r$ 
describe the radial wave functions in the open and closed channels, respectively. 
We denote by $E_o$ and $E_c$ the corresponding threshold energies, defined as the sums of the internal energies of the two atoms in each channel. 
Without loss of generality, we set the zero of energy at the open-channel threshold, $E_o = 0$. 
The closed-channel threshold $E_c$ depends on the internal spin configuration and can be tuned with an external magnetic field through the Zeeman effect, 
which provides experimental control over the relative energy between the two channels.

To keep the model analytically tractable, we assume square-well potentials for the intra- and inter-channel interactions. Here $U_0>0$ and $U_c>0$ are the intrinsic depths of the open- and closed-channel wells, $W>0$ is the coupling strength between the two channels, and $b$ is the range of the interaction. The interactions are taken to be
\begin{equation}
\begin{aligned}
& r < b: \quad
V_{oo}(r) = -U_0, \quad
V_{cc}(r) + E_c(B) = -U_c, \\
& \hspace{3.5cm} V_{oc}(r) = V_{co}(r) = -W; \\
& r > b: \quad
V_{oo}(r) = V_{oc}(r) = V_{co}(r) = 0, \quad
V_{cc}(r) = +\infty.
\end{aligned}
\end{equation} 

The explicit dependence $E_c(B)$ reflects the magnetic-field tunability of the closed-channel threshold, 
while the potential depths $U_0$ and $U_c$ themselves are fixed by the electronic structure of the atoms. 
Although minimal, this model captures the essential features of a Feshbach resonance: 
a near-threshold bound state in the closed channel can couple to the open channel and 
strongly modify its scattering properties, as we will now discuss.

The coupled radial equations corresponding to Eq.~\eqref{lithium:psietwochan} inside the interaction region ($r<b$) take the matrix form
\begin{equation}
\!\!
-\frac{\hbar^2}{2m}\frac{d^2}{dr^2}
\begin{bmatrix}
 y_{E,o} \\[3pt] y_{E,c}
\end{bmatrix}
+
\begin{bmatrix}
 -U_0 & W \\[3pt]
 W & -U_c - E_c(B)
\end{bmatrix}
\!
\begin{bmatrix}
 y_{E,o} \\[3pt] y_{E,c}
\end{bmatrix}
= 
E
\begin{bmatrix}
 y_{E,o} \\[3pt] y_{E,c}
\end{bmatrix}.
\label{lithium:twowelleqinside}
\end{equation}
Outside the interaction region ($r>b$), the closed-channel component vanishes, $y_{E,c}(r)=0$, 
while the open channel has the asymptotic form
\begin{equation}
  y_{E,o}(r) = \sqrt{\frac{2}{\pi\hbar}}\,\sin\!\left(\frac{pr}{\hbar} + \delta_0(p)\right),  
\end{equation}
with $E = p^2/(2m)$. 
Matching these solutions at $r=b$ determines the phase shift $\delta_0(p)$ in the open channel.

The details of the analytical solution are provided in the Supplementary Material.
For compactness, we define an effective closed-channel depth
\begin{equation}
U_c^{\mathrm{eff}}(B) \equiv U_c + E_c(B),
\end{equation}
which combines the intrinsic potential depth $U_c$ with the magnetic-field–dependent 
threshold shift $E_c(B)$.

The $s$-wave scattering length $a_0$ is
\begin{equation}
\frac{1}{a_0 - b}
= -\,\frac{1}{\hbar}\! \lim_{p \to 0}
\left[
\frac{p_{+}\cos^{2}\theta}{\tan\!\big(\tfrac{p_{+}b}{\hbar}\big)}
+
\frac{p_{-}\sin^{2}\theta}{\tan\!\big(\tfrac{p_{-}b}{\hbar}\big)}
\right].
\label{lithium:a0twochannels}
\end{equation}
where $p_{\pm}$ and $\theta$ are defined by
\begin{equation}
\begin{aligned}
& \frac{p_{\pm}^{2}}{2m}
=E + \frac{U_0+U_c^{\mathrm{eff}}(B)}{2}
\pm
\sqrt{\!\left[\frac{U_0-U_c^{\mathrm{eff}}(B)}{2}\right]^2\!+W^2},\\[3pt]
& \sec(2\theta)
=\sqrt{1+\!\left[\frac{2W}{\,U_0-U_c^{\mathrm{eff}}(B)\,}\right]^2}.
\end{aligned}
\label{eq:p_pm_theta_defs}
\end{equation}
The quantities $p_{\pm}$ correspond to the effective momenta associated with the two coupled channels, while the angle $\theta$ parametrizes the mixing between the open and closed channels induced by the coupling $W$. In particular, $\theta\to0$ in the weak-coupling limit $W\to0$, where the two channels become effectively independent.

We comment that, although all the quantities in \eqref{eq:p_pm_theta_defs} are well-behaved in the limit $p\to 0$, divergences can appear due to the denominators in \eqref{lithium:a0twochannels} when very low-energy bound states appear, as we will see below, and so we do not take the limit at this level.

Although Eq.~\eqref{lithium:a0twochannels} looks involved, useful insight comes from special limits.
First, for \emph{no interchannel coupling} ($W=0$) one has $\theta=0$ and $p_{+}=\sqrt{2mU_0}$, giving
\begin{equation}
\frac{1}{a_{0,\mathrm{bg}} - b}
= -\,\frac{1}{\hbar}\,
\frac{\sqrt{2mU_0}}{\tan\!\Big(\tfrac{\sqrt{2mU_0}}{\hbar} b\Big)},
\end{equation}
which is precisely Eq.~\eqref{BasicScatt:scatlengthsqwell}. We will refer to this as the ``background'' scattering length.

Next, consider the \emph{weak-coupling} regime 
$W \ll |U_0-U_c^{\mathrm{eff}}(B)|$, so that $|\theta|\ll1$.
Then Eq.~\eqref{lithium:a0twochannels} reduces to
\begin{equation}
\frac{1}{a_0 - b}
\simeq
\frac{1}{a_{0,\mathrm{bg}} - b}
-\frac{1}{\hbar}\, \lim_{p \to 0}
\frac{p_{-}\,\theta^{2}}{\tan\!\big(\tfrac{p_{-}b}{\hbar}\big)}.
\label{eq:a0_weak_coupling}
\end{equation}

To make contact with the role of bound states (Sec.~\ref{Sec:squarewell}), 
suppose that the closed channel supports a \emph{bound state} with energy 
$E_{\mathrm{bound}}(B)$ (tunable with a magnetic field) very close to threshold ($E_{\mathrm{bound}}(B)\ll U_c^{\mathrm{eff}}(B)$). 
The corresponding radial wave function inside the potential well has the form
\begin{equation}
 y_{\mathrm{bound}}(r)
=\sin\!\left(\frac{\sqrt{2m\,[E_{\mathrm{bound}}(B)+U_c^{\mathrm{eff}}(B)]}}{\hbar}\,r\right),   
\end{equation}
and:
\begin{equation}
\frac{\sqrt{2m\,[E_{\mathrm{bound}}(B)+U_c^{\mathrm{eff}}(B)]}}{\hbar}\,b = n \pi, \quad n \in \mathbb{N}
\end{equation}
The last term in Eq.~\eqref{eq:a0_weak_coupling} (for finite $E$)
can be expanded for $E\!\approx\!E_{\mathrm{bound}}(B)$ as
\begin{equation}
-\frac{1}{\hbar}\,
\frac{p_{-}\,\theta^{2}}{\tan\!\big(\tfrac{p_{-}b}{\hbar}\big)}
\;\approx\;
\frac{2\theta^{2}\,[E_{\mathrm{bound}}(B)+U_c^{\mathrm{eff}}(B)]}
{E-E_{\mathrm{bound}}(B)}.
\label{eq:bound_expansion}
\end{equation}
Substituting this result into Eq.~\eqref{eq:a0_weak_coupling} and taking the limit $E \to 0$ yields
\begin{equation}
\frac{1}{a_0 - b}
=
\frac{1}{a_{0,\mathrm{bg}} - b}
+\frac{\Gamma/2}{E_{\mathrm{bound}}(B)\,b},
\quad
\frac{\Gamma}{2}
\equiv
\frac{2W^{2}\,U_c^{\mathrm{eff}}(B)}
{\big[U_0-U_c^{\mathrm{eff}}(B)\big]^2}.
\label{lithium:a0_poleform}
\end{equation}
Thus, in the weak-coupling limit a near-threshold bound state ($E_{\mathrm{bound}}(B) \to 0^{-}$) in the 
\emph{closed} channel produces a strong enhancement of $a_0$ in the 
\emph{open} channel.

\begin{figure}[t]  % 't' means top of the column; you can also use [h] or [b]
    \centering
    \includegraphics[width=\columnwidth]{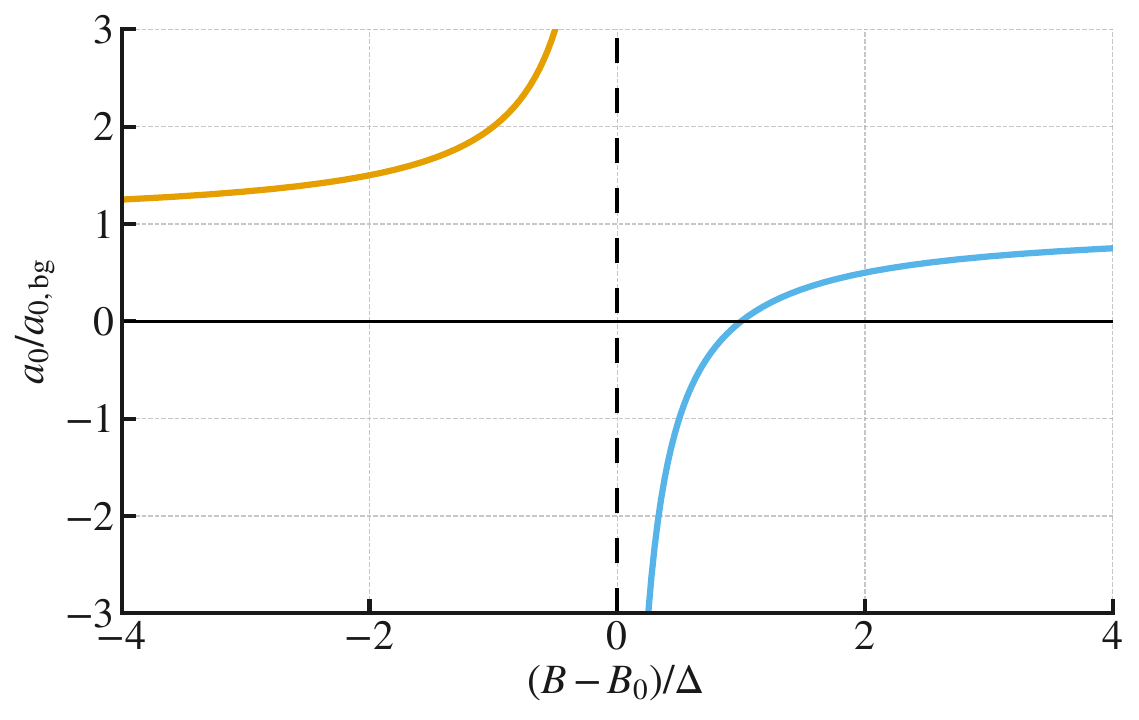}
    \caption{$a_0/a_{0,bg}$ vs. $(B-B_0)/\Delta$ relationship shown in \eqref{lithium:aofB}. We note that a distinct resonance occurs when the external field is tuned to $B=B_0$. The horizontal line corresponds to the scattering length when there is no coupling between the two channels, ``background''.}
    \label{fig: B-B_0/delta plot}
\end{figure}

Because $E_c(B)$ is Zeeman-tunable, the bound-state energy may be taken 
locally linear in $B$,
\begin{equation}
E_{\mathrm{bound}}(B)\approx\mu(B-B_c),
\end{equation}
where $\mu$ is an effective magnetic moment and $B_c$ the field at which 
the closed-channel bound state crosses the open-channel threshold.
Combining this with Eq.~\eqref{lithium:a0_poleform} yields the standard 
Feshbach-resonance expression
\begin{equation}
a_0(B)
=a_{0,\mathrm{bg}}
\!\left(
1-\frac{\Delta}{B-B_0}
\right),
\label{lithium:aofB}
\end{equation}
with width and position
\begin{equation}
\Delta
\equiv
\frac{(a_{0,\mathrm{bg}}-b)^2}{\mu\,a_{0,\mathrm{bg}}\,b}\,
\frac{\Gamma}{2},
\qquad
B_0
\equiv
B_c-
\frac{a_{0,\mathrm{bg}}\Delta}{a_{0,\mathrm{bg}}-b}.
\end{equation}
Equation~\eqref{lithium:aofB}, shown in Fig.~\ref{fig: B-B_0/delta plot}, 
defines a \emph{Feshbach resonance}: by tuning $B$ such that the 
closed-channel bound state aligns with threshold, the scattering length 
diverges. 
This mechanism requires, as a necessary but not sufficient condition, at least one closed channel coupled to the 
entrance channel and relies on the magnetic-field sensitivity of the 
internal (hyperfine/Zeeman) energies entering $U_c^{\mathrm{eff}}(B)$.

Finally, we briefly consider the case of strong coupling, $W \approx |U_0 - U_c|$.
From \eqref{lithium:a0twochannels} we find it convenient to 
define two ``dressed'' channels with effective well depths
\begin{equation}
U^{\mathrm{eff}}_{\pm}(B) = \frac{U_0 + U_c^{\mathrm{eff}}(B)}{2} \pm \sqrt{ \left(\frac{U_0 - U_c^{\mathrm{eff}}(B)}{2}\right)^2 + W^2 }.
\end{equation}
A bound state of the coupled system now corresponds to one of these dressed channels, satisfying
\begin{equation}
\sin \left( \hbar^{-1} \sqrt{2 m \big(E_{\rm bound,\pm}(B) + U^{\mathrm{eff}}_{\pm}(B)\big)} \, b \right) = 0.
\end{equation}
As in the weak-coupling regime, these energies can be tuned with an external magnetic field via the Zeeman effect, e.g.,
\begin{equation}
E_{\rm bound,\pm}(B) \approx \mu_\pm \, (B - B_\pm),
\end{equation}
where $B_\pm$ corresponds to the field at which the dressed state aligns with the open-channel threshold.
From \eqref{eq:p_pm_theta_defs} we can write
\begin{equation}
    \frac{p_{\pm}^2}{2m} = E - E_{\rm bound,\pm}(B) + E_{\rm bound,\pm}(B) + U_{\rm eff,\pm}.
\end{equation}
Now, say we tune $B$ in such a way that $E_{\rm bound,+}(B)$ is close to the open channel threshold (in principle both $E_{\rm bound,\pm}(B)$ can be close to such threshold, but we ignore this possibility for simplicity). We can proceed exactly in the same way as before and write the expansion
\begin{equation}
\begin{split}
    & - \hbar^{-1} \frac{p_{+} \cos^2 \theta}{2 \tan\left(\frac{p_{+}}{\hbar}b  \right)} 
    \approx - \frac{\cos^2 \theta(E_{\rm bound,+}(B) + U_{\rm eff,+}) }{ (E - E_{\rm bound,+}(B)) b } 
     \end{split}
\end{equation}
Some lengthy but simple manipulations, using the definition of $\theta$, allow us to express the right-hand side, in the zero energy limit, as $\frac{\Gamma_{+}/2}{E_{\rm bound,+}(B)b}$, which has exactly the same form we found in the weak coupling limit.
Tuning $B$ shifts the positions of the poles of $a_0$, producing broad Feshbach resonances. Compared to the weak-coupling case, each resonance now originates from a ‘dressed’ bound state, leading to wider and sometimes overlapping resonances. This mechanism allows the scattering properties to be modified over a broad range of magnetic fields, and both narrow (weak-coupling) and broad (strong-coupling) resonances can be observed.

\FloatBarrier

\section{Conclusion}
\label{Sec:conclusion}
Feshbach resonances provide a powerful mechanism for tuning interactions between ultracold atoms or molecules using external magnetic fields. This tunability is especially important because the interaction strength strongly influences the collective behavior of ultracold quantum gases. By controlling the scattering length, experiments can access regimes ranging from weakly interacting gases to strongly correlated many-body systems, enabling the study of phenomena such as superfluidity, molecular formation, and universal quantum behavior. Although these phenomena play a central role in modern atomic physics, their underlying principles are often presented at a level inaccessible to students encountering them for the first time.

In this work, we have developed a systematic and pedagogically oriented framework that connects the study of Feshbach resonances to core topics of undergraduate quantum mechanics. Starting from familiar concepts—scattering theory and spherical potentials—we built up to the description of coupled channels and the emergence of resonances in a two-channel model. This progression highlights how near-threshold bound states and interchannel coupling lead naturally to the magnetic-field–dependent control of scattering properties.

We hope that this presentation will serve as a useful resource for instructors and students, providing both a conceptual foundation and an accessible entry point into the broader and rapidly evolving field of ultracold atomic physics.

\begin{acknowledgments}

This work was supported by the National
Science Foundation Award No. PHY-2207048.

\end{acknowledgments}

\section*{Author Contributions}

Ettore Vitali developed the general approach and presentation of the paper. 
Gino Edward Gamboni researched the relevant literature on lithium hyperfine states and the theory of Feshbach resonances. 
Both authors contributed to the preparation and revision of the manuscript.

%The authors have no conflicts to disclose.

\bibliography{Bibliography}

@article{taron2013,
  author = {Taron, Josep},
  title = {Feshbach resonance: A one dimensional example},
  journal = {American Journal of Physics},
  volume = {81},
  number = {8},
  pages = {603--609},
  year = {2013},
  doi = {10.1119/1.4804193}
}

@article{FESHBACH1958357,
title = {Unified theory of nuclear reactions},
journal = {Annals of Physics},
volume = {5},
number = {4},
pages = {357-390},
year = {1958},
issn = {0003-4916},
doi = {https://doi.org/10.1016/0003-4916(58)90007-1},
url = {https://www.sciencedirect.com/science/article/pii/0003491658900071},
author = {Herman Feshbach}
}

@article{RevModPhys.82.1225,
  title = {Feshbach resonances in ultracold gases},
  author = {Chin, Cheng and Grimm, Rudolf and Julienne, Paul and Tiesinga, Eite},
  journal = {Rev. Mod. Phys.},
  volume = {82},
  issue = {2},
  pages = {1225--1286},
  numpages = {0},
  year = {2010},
  month = {Apr},
  publisher = {American Physical Society},
  doi = {10.1103/RevModPhys.82.1225},
  url = {https://link.aps.org/doi/10.1103/RevModPhys.82.1225}
}

@article{RevModPhys.80.885,
  title = {Many-body physics with ultracold gases},
  author = {Bloch, Immanuel and Dalibard, Jean and Zwerger, Wilhelm},
  journal = {Rev. Mod. Phys.},
  volume = {80},
  issue = {3},
  pages = {885--964},
  numpages = {0},
  year = {2008},
  month = {Jul},
  publisher = {American Physical Society},
  doi = {10.1103/RevModPhys.80.885},
  url = {https://link.aps.org/doi/10.1103/RevModPhys.80.885}
}

@book{griffiths,
	address = {Cambridge ; New York, NY},
	edition = {Third edition},
	title = {Introduction to quantum mechanics},
	isbn = {978-1-107-18963-8},
	publisher = {Cambridge University Press},
	author = {Griffiths, David J. and Schroeter, Darrell F.},
	year = {2018},
}

@article{
doi:10.1126/science.1079107,
author = {K. M. O'Hara  and S. L. Hemmer  and M. E. Gehm  and S. R. Granade  and J. E. Thomas },
title = {Observation of a Strongly Interacting Degenerate Fermi Gas of Atoms},
journal = {Science},
volume = {298},
number = {5601},
pages = {2179-2182},
year = {2002},
doi = {10.1126/science.1079107},
URL = {https://www.science.org/doi/abs/10.1126/science.1079107},
eprint = {https://www.science.org/doi/pdf/10.1126/science.1079107}}

@article{PhysRevLett.91.080406,
  title = {Conversion of an Atomic Fermi Gas to a Long-Lived Molecular Bose Gas},
  author = {Strecker, Kevin E. and Partridge, Guthrie B. and Hulet, Randall G.},
  journal = {Phys. Rev. Lett.},
  volume = {91},
  issue = {8},
  pages = {080406},
  numpages = {4},
  year = {2003},
  month = {Aug},
  publisher = {American Physical Society},
  doi = {10.1103/PhysRevLett.91.080406},
  url = {https://link.aps.org/doi/10.1103/PhysRevLett.91.080406}
}

@book{joachain1975quantum,
  author    = {Joachain, C.~J.},
  title     = {Quantum Collision Theory},
  publisher = {North--Holland Publishing Co. / American Elsevier},
  address   = {Amsterdam / New York},
  year      = {1975},
  edition   = {1st},
  isbn      = {0-7204-0294-8},
}

@book{landau_lifshitz_qm,
  author       = {Landau, L.~D. and Lifshitz, E.~M.},
  title        = {Quantum Mechanics (Non‑relativistic Theory)},
  series       = {Course of Theoretical Physics},
  volume       = {3},
  publisher    = {Pergamon Press},
  address      = {Oxford},
  year         = {1977},
  edition      = {3rd},
  isbn         = {0-08-021017-1},
  note         = {See §34 for derivation of plane‑wave expansion coefficients},
}

@article{fetic2024wigner,
  author       = {Fetić, Benjamin and Becker, Wilhelm and Milošević, Dejan B.},
  title        = {Wigner time delay revisited},
  journal      = {Annals of Physics},
  volume       = {465},
  pages        = {169666},
  year         = {2024},
  doi          = {10.1016/j.aop.2024.169666},
  note         = {“The concept of Wigner time delay was introduced in scattering theory to quantify the delay or advance of an incoming particle in its interaction with the scattering potential.”},
}

@article{griffiths1982hyperfine,
  author    = {Griffiths, David J.},
  title     = {Hyperfine splitting in the ground state of hydrogen},
  journal   = {American Journal of Physics},
  volume    = {50},
  number    = {8},
  pages     = {698--702},
  year      = {1982},
  doi       = {10.1119/1.12733},
  note      = {“The hyperfine structure of atomic hydrogen is derived in a simple and self‑contained way that makes the theory accessible to advanced undergraduates.”},
}

@article{schlecht1966li_hyperfine,
  author    = {Schlecht, Richard G. and McColm, Douglas W.},
  title     = {Hyperfine Structure of the Stable Lithium Isotopes.I},
  journal   = {Physical Review},
  volume    = {142},
  number    = {1},
  pages     = {11--17},
  year      = {1966},
  doi       = {10.1103/PhysRev.142.11},
  note      = {Reports $\Delta\nu_6$=228.20528(8) Mc/s for $^6$Li hyperfine splitting},
}

@article{houbiers1997li_collisions,
  author    = {Houbiers, M. and Stoof, H.~T.~C. and McAlexander, W.~I. and Hulet, R.~G.},
  title     = {Elastic and inelastic collisions of $^6$Li in magnetic and optical traps},
  journal   = {Physical Review A},
  volume    = {56},
  number    = {6},
  pages     = {4864--4870},
  year      = {1997},
  doi       = {10.1103/PhysRevA.56.4864},
  note      = {Uses full coupled‑channel method for ultracold $^6$Li collision properties},
}

@book{eisberg_resnick_1985,
  author    = {Eisberg, Robert and Resnick, Robert},
  title     = {Quantum Physics of Atoms, Molecules, Solids, Nuclei, and Particles},
  edition   = {2nd},
  publisher = {Wiley},
  year      = {1985},
  isbn      = {9780471873735},
  note      = {Covers the Zeeman effect in atomic spectra, see Chapter 7},
}

@book{griffiths_qm_2018_zeeman,
  author    = {Griffiths, David J. and Schroeter, Darrell F.},
  title     = {Introduction to Quantum Mechanics},
  edition   = {3rd},
  publisher = {Cambridge University Press},
  year      = {2018},
  isbn      = {9781107189638},
  note      = {Chapter 6.2 discusses the Zeeman effect in the context of perturbation theory},
}

\end{document}